\documentclass[onecolumn,notitlepage,nofootinbib,longbibliography,11pt]{revtex4-1}
\usepackage{multirow}
\usepackage{amssymb, esvect, amsmath, graphicx, latexsym, amsthm, slashed, eso-pic}
\usepackage{xcolor}
\usepackage{amsmath,amsthm,amsfonts,amscd,amssymb} 
\usepackage{eucal}
\usepackage{braket}
\usepackage{slashed}
\usepackage{hyperref}
\usepackage{graphicx}
\usepackage{bm}
\usepackage{xcolor}
\usepackage{afterpage}
\usepackage{changepage}

\usepackage{hyperref}
\DeclareMathOperator{\ev}{eV}

   \def\oL{\overline} 
       
\newcommand{\beq}{\begin{equation}} \newcommand{\eeq}{\end{equation}}
\newcommand{\bea}{\begin{eqnarray}} \newcommand{\eea}{\end{eqnarray}}

\newcommand{\Li}{\mathcal{L}}
\newcommand{\Oi}{\mathcal{O}}

\def\lsim{\mathrel{\raise.3ex\hbox{$<$\kern-.75em\lower1ex\hbox{$\sim$}}}}
\def\gsim{\mathrel{\raise.3ex\hbox{$>$\kern-.75em\lower1ex\hbox{$\sim$}}}}

\newcommand{\be}{\begin{eqnarray}}
\newcommand{\ee}{\end{eqnarray}}

\newcommand{\benum}{\begin{enumerate}}
\newcommand{\eenum}{\end{enumerate}}
\newcommand{\bi}{\begin{itemize}}
\newcommand{\ei}{\end{itemize}}

\begin{document}

\preprint{FERMILAB-PUB-19-364-T}

\title{Ultralight dark matter detection with mechanical quantum sensors}

\author{Daniel Carney$^{a,b}$}
\author{Anson Hook$^{c}$}
\author{Zhen Liu$^c$}
\author{Jacob M. Taylor$^{a}$}
\author{Yue Zhao$^{d}$}

\medskip

\affiliation{$^a$Joint Quantum Institute/Joint Center for Quantum Information and Computer Science, University of Maryland, College Park/National Institute of Standards and Technology, Gaithersburg, MD, USA}
\affiliation{$^b$Fermi National Accelerator Laboratory, Batavia, IL, USA}
\affiliation{$^c$Maryland Center for Fundamanetal Physics, University of Maryland, College Park, MD USA}
\affiliation{$^d$Department of Physics and Astronomy, University of Utah, Salt Lake City, UT 84112, USA}

\date{\today}

\begin{abstract}
We consider the use of quantum-limited mechanical force sensors to detect ultralight (sub-meV) dark matter candidates which are weakly coupled to the standard model. We show that mechanical sensors with masses around or below the milligram scale, operating around the standard quantum limit, would enable novel searches for dark matter with natural frequencies around the kHz scale.  This would complement existing strategies based on torsion balances, atom interferometers, and atomic clock systems.
\end{abstract}

\maketitle

\tableofcontents

\section{Introduction}

There is overwhelming astrophysical and cosmological evidence for the existence of cold dark matter (DM) \cite{Sofue:2000jx,Massey:2010hh,Aghanim:2018eyx,Primack:2015kpa}. However, the detailed properties of the DM, in particular its mass, are highly unconstrained \cite{Tanabashi:2018oca}. The possibility that the dark sector could contain or even consist primarily of ``ultralight'' bosonic fields of mass $10^{-22}~{\rm eV} \lesssim m_{\phi} \lesssim 0.1~{\rm eV}$ has recently received considerable attention~\cite{Essig:2013lka,Hui:2016ltb,Irastorza:2018dyq}.

The existence of such light fields coupling to the standard model (SM) may induce new forces beyond the usual electroweak and strong forces. For example, if there are additional light scalars which couple to SM matter through Yukawa couplings, the best constraints come from torsion balance experiments looking for ``5th forces'' between a pair of macroscopic masses \cite{adelberger2009torsion,wagner2012torsion}. Furthermore, as first emphasized in \cite{graham2016dark}, if these ultralight bosons make up a significant fraction of the ambient DM background, existing mechanical accelerometers \cite{adelberger2009torsion,wagner2012torsion,pierce2018searching,Guo:2019ker} and upcoming large-scale atomic interferometers (eg. \cite{Coleman:2018ozp}) can be used to further probe these models. 

Ultralight dark matter detection can be viewed as a paradigmatic exercise in quantum metrology. Essentially, the dark matter creates a classical, persisent, but extremely weakly-coupled force on a sensor. The goal is to then detect or rule out the presence of this force. This task is fundamentally limited by the noise in the detection system. High-quality microwave cavity systems are already used for such metrological searches for a particular ultralight DM candidate--the axion, which couples to the electromagnetic field in the cavity \cite{du2018search,zhong2018results}. Here we instead focus on ultralight DM candidates which couple to all atoms in a massive object coherently, for example through direct coupling to neutron number.

In this paper, we study the prospects for searching for these ultralight DM candidates using massive, mechanical sensors. These devices consist of optical or microwave fields optomechanically coupled to a diverse range of mechanical systems, including suspended pendulums \cite{TheLIGOScientific:2014jea,matsumoto20155,matsumoto2019demonstration} and torsion balances \cite{adelberger2009torsion,wagner2012torsion}, high-tension mechanical membranes \cite{reinhardt2016ultralow,tsaturyan2017ultracoherent}, levitated dielectric \cite{yin2013optomechanics,chang2010cavity,hempston2017force,monteiro2017optical} or magnetic \cite{chan1987superconducting1,chan1987superconducting2,prat2017ultrasensitive} objects, and even levitated liquids \cite{childress2017cavity}. Our primary goal is to understand the parametric scalings in sensitivity with a variety of systems of different masses, frequencies, and noise levels, rather than to focus on a specific experimental setup. This complements the detailed analysis based on the E\"{o}tv\"{o}s-Washington experiments presented in \cite{graham2016dark}. 

We emphasize the utility of multiple sensing devices (in an array or distributed at widely separated locations) for DM detection. Arrays of sensors benefit from reduction of the noise compared to the correlated DM signal, and enable rejection of some important backgrounds. In particular, the signals we consider here act differently on different material types, unlike the gravitational and seismic noise backgrounds dominating at low frequencies. Thus an array with at least two different types of material sensors can be used to make differential measurements which subtract out these backgrounds \cite{graham2016dark}. We will see that at the level of a pair of sensors in a well-isolated environment (eg. dilution refrigeration) these mechanical sensors already have detection reach competitive with existing torsion balance and interferometery experiments. Arrays of $N_{\rm det}$ sensors can further enhance the detection reach by a factor of at least $\sqrt{N_{\rm det}}$, and as fast as $N_{\rm det}$, in the case of correlated, ``Heisenberg-limited'' readout \cite{giovannetti2011advances}. We find that optomechanical systems are well-poised to contribute to searches for ultralight dark matter for masses below around $10^{-8}~{\rm eV}$; for higher masses the laser power requirements in the readout become very stringent (see figure~\ref{figure-limits}). Detection frequencies in the kHz-MHz range appear to be the most promising.

Recently, it has been proposed that a large array of mechanical sensors operating with significant quantum noise-reduction could enable direct detection of heavy ($m \gtrsim m_{\rm pl}$) DM candidates purely through their gravitational interactions \cite{Carney:2019pza}. Here, we propose ultralight DM detection as a nearer-term goal achievable with just a few mechanical sensors, operating at noise levels around the ``standard quantum limit'' (SQL), a benchmark already demonstrated in many devices. The scalable nature of an array and potential improvements to the noise levels beyond the SQL would allow for significant ultralight DM detection reach en route 
to the long-term goal of gravitational DM detection with mechanical sensors.

\section{Ultralight dark matter}

\begin{figure}[t]
\includegraphics[scale=1.1]{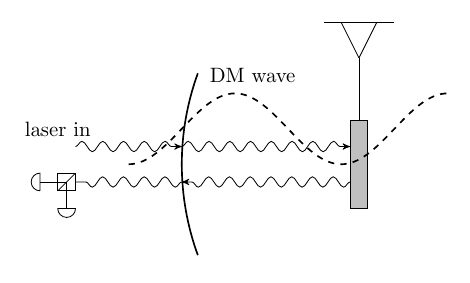}
\caption{Schematic of a single-sided cavity optomechanics search for ultralight DM. Photons reflected off the suspended mirror pick up phases proportional to the mirror's position $x(t)$. This information is then read out via standard interferometry. The ultralight DM produces an essentially monochromatic force with wavelength much longer than the size of the sensor; the presence of this force can then be inferred by reading out the cavity light. In this setup with a single sensor, the fixed reference mirror and movable sensor mirror should be made from different materials, as discussed near equation \eqref{forcegeneral}.}
\label{cavity-opto}
\end{figure}

To begin, we briefly review the salient facts about ultralight dark matter. The essential feature of these ultralight candidates is that they behave as a persistent, wave-like field, due to their high occupation number. This should be contrasted to heavier DM candidates, for example ${\rm GeV}$-scale weakly interacting massive particles (WIMPs), which manifest as a dilute gas of single particles. See \cite{Lisanti:2016jxe} for a general review of dark matter properties, and \cite{Arias:2012az,Irastorza:2018dyq} for reviews of ultralight dark matter.

Suppose the dark sector consists of a single bosonic field of mass $m_{\phi}$. Bosonic dark matter is required to have $m_{\phi} \gtrsim 10^{-22} ~\ev$ so that its de Broglie wavelength does not exceed the core size of a dwarf galaxy and to satisfy Lyman-$\alpha$ constraints~\cite{Hui:2016ltb,Armengaud:2017nkf,Marsh:2018zyw,Bar:2019bqz} (note that recent evidence indicates that the constraints may be a few orders of magnitude stronger than this~\cite{Bar:2018acw,Bar:2019bqz,Safarzadeh:2019sre}). Dark matter this light is necessarily bosonic as fermionic dark matter with a mass $\lesssim$ keV would not fit inside of a dwarf galaxy due to the Pauli exclusion principle~\cite{Tremaine:1979we}. Assuming that the DM has virialized to the galaxy, it will be moving with a typical speed $v \sim 10^{5} ~{\rm m/s}$ due to the viral theorem, and thus have de Broglie wavelength of order $\lambda = 1/m_{\phi} v$.\footnote{In this paper we use natural units $\hbar = c = 1$ in equations, but will quote experimentally-relevant numbers in SI units for ease of comparison with experimental work.}

\begin{figure}[t]
    \centering
    \includegraphics[width=0.46\linewidth]{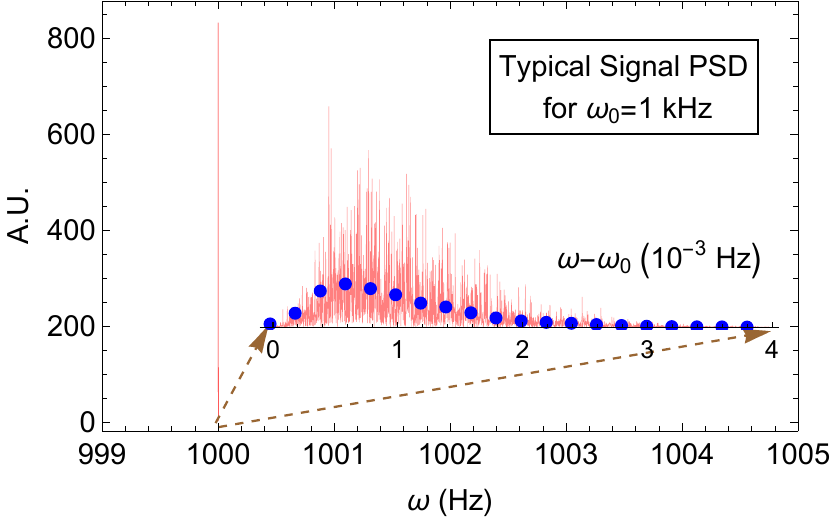}
    \includegraphics[width=0.46\linewidth]{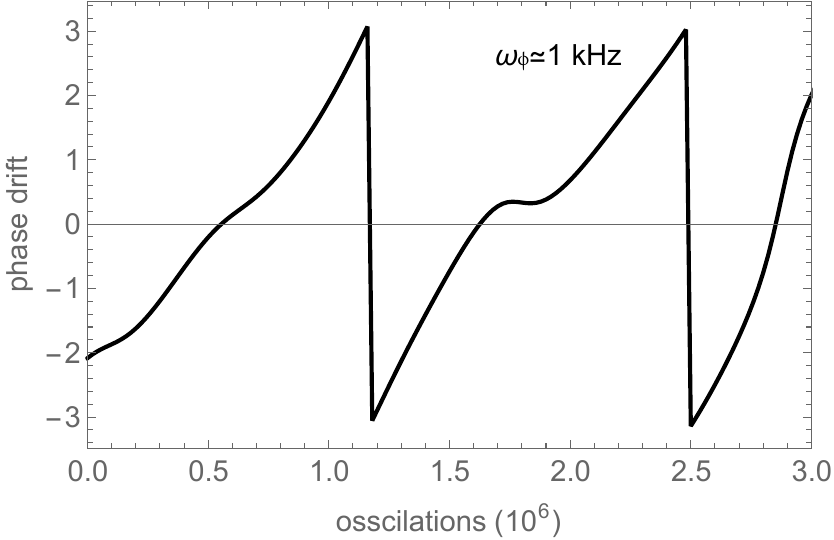}
    \caption{Example simulated power spectral density and phase drift for vector dark matter of mass $m_{\phi} =6.6\times 10^{-13}~{\rm eV}$, corresponding to a signal frequency $\omega_0 = 1~{\rm kHz}$ at zero momentum. One can read off the effective quality factor of the DM signal $Q \sim 10^6$, which sets the coherence time of the signal. In terms of the force signal expected on a device, the left plot represents precisely the shape of the force signal PSD, with the scale set by the coefficient in \eqref{forcenodelta}. See appendix \ref{app-DMPSD} for details.}
    \label{fig:signal}
\end{figure}

Consider the number of DM quanta in a typical cell of phase space. If $n_{\rm DM} = \rho_{\rm DM}/m_{\phi}$ is the number density of DM, the phase space occupancy is given by
\be
\lambda^3 n_{\rm DM} = \frac{\rho_{\rm DM}}{m_{\phi}^4 v^3} \sim 10^{15} \left( \frac{1~{\rm meV}}{m_{\phi}} \right)^4,
\ee
where the observed value of the local dark matter density $\rho_{\rm DM} \sim 0.3~{\rm GeV}/{\rm cm}^3$. Thus for any $m_{\phi} \lesssim 0.1~\ev$, this occupancy is huge, indicating that the dark matter can be treated as a classical field, essentially a superposition of many different plane waves. These waves have velocities following a Boltzmann distribution. The phases in these plane waves are uncorrelated from each other.  Their propagation directions (and polarization vectors, for vectorial fields) are distributed isotropically as long as DM has fully virialized.  Superposing a large number of such waves will produce an overall time-dependent waveform with frequency $\omega \simeq m_{\phi}$ and frequency spread $\delta \omega \approx v/\lambda = v^2 m_{\phi}$. This frequency spread leads to a coherence time $T_{\rm coh} \approx 1/\delta \omega \approx 10^6/\omega$. For a time duration smaller than $T_{\rm coh}$, the DM background field can be approximately treated as a coherent sinusoidal wave with signal angular frequency and wavelength
\be
\omega_{\phi} \simeq 10^{12} ~ {\rm Hz} \times \left( \frac{m_{\phi}}{1~{\rm meV}} \right), \ \lambda \simeq 1~{\rm m} \times \left( \frac{1~{\rm meV}}{m_{\phi}} \right).
\label{freqwavelength}
\ee
In figure \ref{fig:signal}, we give a visualization of the power spectral density of the DM field, generated by simulating a large number of superposed DM quanta (see appendix \ref{app-DMPSD} for details). In particular, one can see the ``quality factor'' of the DM signal $Q \sim 10^6$ coming from the frequency drift $\delta \omega$.

Assuming there is a non-zero coupling between the DM field and the sensor, if the observational time $T_{\rm int}$ is much shorter than $T_{\rm coh}$,  we can parametrize the force exerted on the sensor (along a given axis) by the DM as
\be
\label{forcenodelta}
F_s(t) = g N_g F_0 \sin(\omega_{\phi} t)
\ee
under the assumption that the wavelength \eqref{freqwavelength} is much larger than the size of the sensor. In equation \eqref{forcenodelta}, $g$ is a dimensionless coupling strength depending on DM model, $F_0 = \sqrt{\rho_{\rm DM}} \sim 10^{-15}~{\rm N}$, and $N_g$ is the number of total charges in a given sensor. For example, consider the case that the light boson is a vector boson which couples to neutrons (studied in detailed in the next subsection); then $N_g \approx \frac{1}{2} m_s/m_{\rm neutron}$ is approximately the number of neutrons in a sensor of mass $m_s$. For observational times longer than $T_{\rm coh}$, we essentially have $N_{\rm bins} = T_{\rm int}/T_{\rm coh}$ independent realizations of \eqref{forcenodelta}, with randomly oriented amplitude and phase. 

To get a sense of the force scales involved, consider \eqref{forcenodelta} with $g \sim 10^{-22}$ (see figure~\ref{figure-limits}). This would generate a force per neutron of around $10^{-38}~{\rm N}$. This would then be detectable with a second of integration time, using a $\mu$g-scale device with force sensitivity in the zepto-Newton regime $10^{-21}~{\rm N}/\sqrt{\rm Hz}$, for example \cite{ranjit2016zeptonewton}.

The force \eqref{forcenodelta} is proportional to the mass of the sensor, and is thus somewhat analogous to a gravitational force. However, crucially, the constant of proportionality is material-dependent. In the neutron-coupled example, two materials with different neutron-to-nucleon ratios will experience a \emph{differential} force. This is a general feature of these types of dark matter--they generate ``equivalence-principle violating'' forces, as emphasized in \cite{graham2016dark}. If the force had satisfied the equivalence principle, then the mechanical element of a sensor and whatever its position is referenced to would both experience the same force, and no motion would be detected (one would have to detect the gradient of the force, i.e.\ the tidal force, see for example \cite{pierce2018searching}). However, these EP-violating forces can be detected directly using a sensor and reference (possibly a second sensor) with different material properties. For example, if the neutron-to-nucleon number ratio of each material is $Z_1/A_1, Z_2/A_2$, then the differential acceleration on the two objects is given by \cite{graham2016dark}
\be
\label{forcegeneral}
a_s(t) = g \Delta F_0 \sin(\omega_{\phi} t), \ \ \ \Delta = \frac{Z_1}{A_1}-\frac{Z_2}{A_2}.
\ee
For typical lab materials, $\Delta \lesssim 0.1$.

Aside from the general properties just mentioned, there are several distinguishing features of the signal that can be used to distinguish from background.  The first is that the signal has a uniform direction over timescales of the coherence time.  This means that force sensors in different directions can be used to isolate the signal while rejecting background.  The second property is that the Earth's motion in the galaxy and self rotation select a preferred direction for signal.  More explicitly, the Earth is moving at speed $v \sim 10^5 ~{\rm m/s}$ around the center of the galaxy.  Additionally, there is increasing evidence that there is a stream of non-virialized DM moving past the Earth with velocities also $v \sim 10^{5} ~{\rm m/s}$~\cite{myeong2017halo,o2018dark,necib2018under}. Thus while it is difficult to say for certain what direction any preference would be, it is certain that the signal will be $\mathcal{O}(1)$ biased in some direction.  The third is that there will be modulations.  The Earth is moving around the sun at $\delta v \sim 10^{4} ~{\rm m/s}$ and the surface of the Earth rotates around its axis as a speed $\delta v \sim 100~{\rm m/s}$.  Thus the DM signal will move fractionally in frequency space by an amount $v \delta v \sim 10^{-7} - 10^{-9}$.\footnote{This comes from the fact that dark matter is non-relativistic so the main effect of adding velocities is to change the energy by $\frac{1}{2} m_{\phi} (v + \delta v)^2$, giving a Doppler shift  $\sim m_{\phi} v \delta v$.} The fourth is that if $m_{\phi} \lesssim 10^{-9}$ eV, the coherence length of DM reaches the size of the Earth, so experiments on multiple parts of the planet would be correlated.

Our general parametrization \eqref{forcegeneral} can be used to model the signal produced by a variety of DM candidates to produce a basic sensitivity estimate. Here, we briefly exhibit a pair of DM models to show how this type of signal arises.

\subsection{Vector B-L dark matter} 

The first case we consider is when dark matter is a spin-1 particle, which couples to a conserved current. In the Standard Model, there are two conserved currents that a vector-like DM can couple to without introducing new gauge anomalies: electric charge, and baryon number minus lepton number, i.e. $B-L$ charge. For the case of electric charge, DM induced force is highly suppressed since the test objects are generically charge neutral, thus we consider the case where dark matter is a vector field that couples to $B-L$ charge.

The Lagrangian in this case can be written as
\be
\Li = - \frac{1}{4} F^2 - \frac{1}{2}m_{\phi}^2 A^2+  i g_{B-L} A_\mu \oL{n}\gamma^\mu n.
\ee
Here $n$ is the neutron field, so the vector boson couples directly to the number of neutrons. The boson mass term can be generated by Higgsing the $U(1)_{B-L}$ group, providing the mass $m_{\phi}$.  For simplicity, the Higgs boson is assumed to be much heavier than the gauge boson, i.e. the Stueckelberg limit. The vector gauge boson is assumed to couple to $B - L$ charge.   For a charge neutral test object, $A_\mu$ effectively couples to the total neutron number. 

Taking the Lorentz gauge, i.e. $\partial^\mu A_\mu = 0$, one can show that $A_0$ is smaller than $A_i$ by a factor of the velocity of dark matter, $v \sim 10^{-3}$.
One also finds that the dark electric field is much larger than the dark magnetic field, i.e. $|\vec E| \simeq |\partial_0 \vec A |\simeq m_{\phi} |\vec A|\gg |\vec B| \simeq m_{\phi} |\vec v| |\vec A|$. In the plane wave approximation, the dark electric field can be written as
\be
E \simeq \sqrt{2 \rho_{\rm DM}} \sin(\omega_{\phi} t - \vec k\cdot\vec x+\phi_0)
\ee
where the normalization is set to reproduce the observed local dark matter mass density, and the frequency is set by the mass of the DM, i.e. $\omega_{\phi} \simeq m_{\phi}$. This will produce a force of the form
\be
\label{BLforce}
F = N_{B-L} g_{B-L} E \approx g_{B-L} N_{B-L} F_0 \sin (\omega_{\phi} t+\phi_0)
\ee
on each sensor.  For the frequencies under consideration, the wavelength is much larger than the size of the experiment so that we will drop the $x$ dependence.  Here for a sensor with mass $m_s$, $N_g = N_{B-L} \simeq \frac{1}{2} m_s/m_{\rm neutron}$, which is the number of neutrons in the sensor and $F_0 = \sqrt{\rho_{\rm DM}} \approx 10^{-15}~{\rm N}$.

\subsection{Scalar coupling to neutrons}

Another option for light DM is that it is a spin-0 particle.  It can couple to SM fermions through two types of couplings, derivative and non-derivative interactions.  Derivative interactions couples to a vector quantity of the test mass.  The test mass tends to have very low intrinsic spin and thus do not couple very strongly to derivative interactions.  As such we consider non-derivative interactions.

The leading non-derivative interactions a scalar can have with the SM are Yukawa interactions, much like the well known example of the Higgs boson.  Similar to the Higgs boson, in the presence of a coherent background of the light scalar field, these Yukawa interactions have the effect of modulating the mass of a particle.  The spatial dependence of the mass results in a force on the test mass.

In order to make a close analogue to the previous example, we imagine a model where the scalar DM $\phi$ couples only to neutrons through a Yukawa coupling
\be
L = -\frac{1}{2}\partial_\mu \phi \partial^\mu \phi-\frac{1}{2} m^2 \phi^2 + y \phi \oL{n} n.
\ee
Under the planewave approximation, the force on an object with $N_g$ neutrons can be written as
\be
F = - \frac{dV}{dx} \approx y N_g v F_0 \sin(\omega_{\phi} t).
\ee
Note that this example is similar to \eqref{BLforce} except that the two couplings are named differently, $g_{B-L}$ versus $y$.  The only other difference comes from the velocity suppression; in terms of the parametrization \eqref{forcegeneral} here we have $g = v y$. This is due to the fact that forces are vector quantities and for the case of a scalar, the only vector quantity to provide a direction is the small velocity.

\section{Detection with optomechanical force sensors}

In this section we give a brief overview of continuous force sensing, with the goal of explaining how force-sensitivity curves are derived in a typical quantum optomechanics experiment. See appendix \ref{optoappendix} for a detailed treatment, or \cite{clerk2010introduction} for a review.

Consider a prototypical force sensor consisting of a high-finesse optical cavity formed by a partially transparent mirror on one side and a high-reflectivity mirror suspended as a pendulum of mass $m_s$ and mechanical frequency $\omega_s$ on the other side (see figure \ref{cavity-opto}). The suspended mirror is used as the sensor, which can be monitored by light sent into the cavity through an external laser. Displacements of the mirror $x(t)$ produce changes of the cavity fundamental frequency, leading to a coupling of the light and mechanics. Light emerging from the cavity picks up a phase dependent on the mechanical displacement, which can then be read out via standard homodyne interferometry. Given the observed displacement signal $x(t)$, we can infer the presence of forces $F(t)$ on the sensor using standard linear response theory.

The classic example of such an optomechanical setup is LIGO \cite{TheLIGOScientific:2014jea}. Here we are describing a simplified version of one of the interferometer arms. The use of mirrors and light is non-essential, however: numerous variants of this idea operate in essentially the same way. Examples include microwave-domain electromechanical systems \cite{reinhardt2016ultralow,tsaturyan2017ultracoherent}, optically-probed levitated objects \cite{yin2013optomechanics,chang2010cavity,hempston2017force,monteiro2017optical}, magnetically-probed and levitated superconducting objects \cite{chan1987superconducting1,chan1987superconducting2,prat2017ultrasensitive}, and even levitated liquids \cite{childress2017cavity}. See \cite{blencowe2004quantum,aspelmeyer2014cavity} for reviews of these types of systems.

The sensitivity of a given sensor is set by noise. In general, the force acting on the sensor is
\be
F(t) = F_{sig}(t) + F_{noise}(t)
\ee
where $F_{sig}$ represents the signal we are looking for and $F_{noise}$ is a random force coming from a variety of noise sources. Assuming that the noise is a stationary random variable, the size of the noise is characterized by the noise force power spectral density (PSD) $S_{FF}$, defined by
\be
\braket{ F_{noise}(\omega) F_{noise}(\omega')} = S_{FF}(\omega) \delta(\omega+\omega').
\ee
The noise force PSD has units of $\text{force}^2/\text{frequency}$. The interpretation of this quantity is that if one wants to detect a signal force of pure frequency $\omega_{\phi}$ and amplitude $F_*$ at the 1-$\sigma$ level, one needs to continuously monitor the sensor for an integration time $T_{\rm int}$ given by 
\be
\label{sensitivity}
F_* = \sqrt{S_{FF}(\omega_{\phi})/T_{\rm int}}.
\ee
The smaller the signal one is looking for, the longer integration time $T_{\rm int}$ is needed.\footnote{Here we assumed the signal is coherent for the entire observation time $T_{\rm int}$. For higher-frequency DM candidates, this may not be satisfied. For details of the scaling behavior, see the discussion in section~\ref{bounds}.} Sensitivities are usually quoted as the square root of $S_{FF}$, say in units of ${\rm N}/\sqrt{{\rm Hz}}$. The appearance of the square root of frequency reflects the Brownian (stationary) character of the noise. 

A number of sources contribute to the noise on a sensor. The two key noise sources we will consider in this paper are thermal noise coming from the coupling of the sensor to its ambient environment as well as measurement-added noise. This latter noise is a fundamental limitation imposed by quantum mechanics: the act of measurement itself induces noise. We write
\be
S_{FF} = S_{FF}^{T} + S_{FF}^{M}.
\ee
In what follows, we will study the fundamental limits to detection of ultralight DM imposed by thermal and measurement noise. A key assumption here is the elimination of correlated technical noise sources, such as seismic noise. Methods for achieving this are discussed in section \ref{arrays}, see also the appendix \ref{optoappendix} for some quantitative details.

\subsection{Thermal noise}

Thermal noise here refers to a random Brownian force $F_{T}(t)$ acting on an individual sensor from thermal fluctuations coupling to the environment. In general, this will come from both direct vibrational coupling of the sensor to its support structure as well as residual gas pressure in the sensor. At the frequencies of interest in this paper, these are white (frequency-independent) stationary noise sources. Simple dimensional analysis (see also appendix \ref{optoappendix}) gives
\be
\label{noise-thermal}
S_{FF}^{T} = \gamma m_s k T + P A_s \sqrt{m_a k T}.
\ee
Here the first term reflects coupling of the sensor to phonons in its support structure; $\gamma$ is the mechanical damping rate of the sensor and $T$ is the temperature of the environment. The second term comes from residual gas pressure in the environment, where $P$ is the gas pressure, $A_s$ is the surface area of the sensor, and $m_a$ is the mass of an individual gas molecule.

\subsection{Measurement noise and the ``standard quantum limit''}
\label{measurementnoise}

Measurement-added noise comes from the quantum mechanics of the act of measurement itself. The measurement noise PSD can be further decomposed into a pair of terms known as shot noise and backaction noise, 
\be
S_{FF}^{M} = S_{FF}^{BA} + S_{FF}^{SN}.
\ee 
In the cavity model given above, backaction noise corresponds to random fluctuations in the input laser amplitude, which causes a random force on the mirror, which is then transduced into the output light. More fundamentally, backaction noise arises as a consequence of the sensor state being measured. Shot noise corresponds to random fluctuations in the input laser phase which then produce a noisy readout.\footnote{Note that shot noise is not an actual force on the mechanical sensor. However, when we use the output light phase to infer force on the sensor, the shot noise contributes an effective noise force in the sense that it will lead to fluctuations in the light intensity readout that propagates to the force interpretation.}

\begin{figure}[t]
    \includegraphics[width=0.49\linewidth]{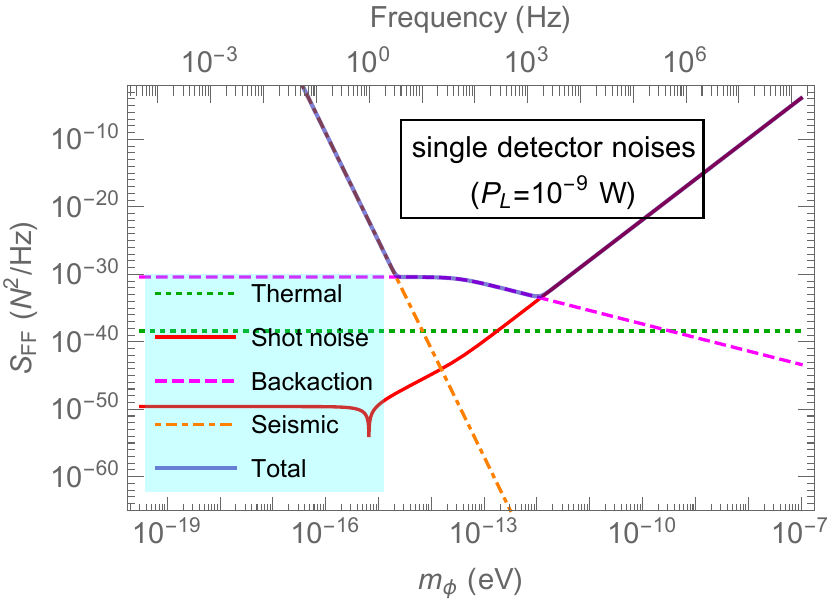}
    \includegraphics[width=0.49\linewidth]{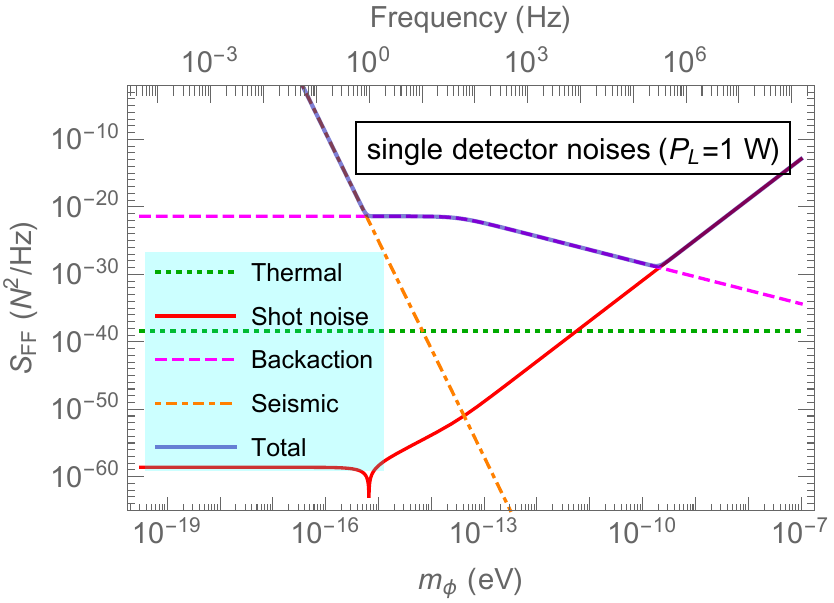}
    \includegraphics[width=0.49\linewidth]{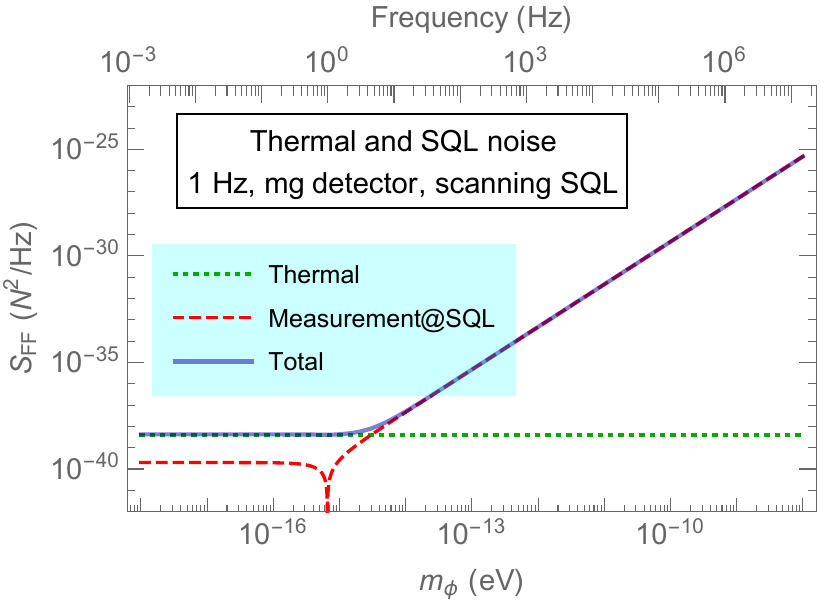}
    \caption{Various contributions to the noise power spectral density of a single sensor, with different input laser powers. Here we have plotted sensitivities for a mechanically-coupled sensor of mass $m_s = 1~{\rm mg}$, mechanical frequency $\omega_s = 1~{\rm Hz}$, and mechanical damping coefficient $\gamma = 10^{-6} ~ {\rm Hz}$ in a $1~{\rm cm}$ optical cavity (parameters based on the setup in \cite{matsumoto2019demonstration}). We assume operation at dilution refrigeration temperatures $T = 10~{\rm mK}$. The incident laser power in the first row is taken as $P_L = 1~{\rm W}$ and $10^{-9}~{\rm W}$ in the left panel and right panel, respectively. The SQL is achieved at the minimum of the total noise curve; the location changes as a function of laser power. In the second row, we show the sensitivity achieved by tuning the laser power to achieve the SQL at each signal frequency $\omega_{\phi}$, for the same sensor parameters (dropping seismic noise for visual simplicity). We see that for these sensor parameters, seismic and thermal noise dominate at low frequencies while measurement-added noise dominates at high frequencies.}
    \label{figure-noisecontribs}
\end{figure}

The backaction and shot noise contributions to the noise PSD are non-trivially dependent on both frequency and readout laser power $P_L$. Backaction tends to dominate at low frequencies while shot noise dominates at high frequencies. Furthermore, as one turns up the laser power, backaction noise increases whereas shot noise decreases. See figure \ref{figure-noisecontribs} for an example of these noise contributions with a typical mechanical sensor.

Given a target signal frequency $\omega_{\phi}$ that one is looking for, the laser power $P_L$ can be tuned to minimize the sum of the backaction and shot noise terms at the frequency $\omega_{\phi}$. This procedure yields the noise at what is known as the ``standard quantum limit'' (SQL), at frequency $\omega_{\phi}$. See figure \ref{figure-noisecontribs} for SQL-level force sensitivities. Achieving the SQL at can require substantial laser power at high frequency, see figure~\ref{figure-laserreqs}, which becomes prohibitive for dark matter masses $m \gtrsim 10^{-8}~{\rm eV}$. As discussed in appendix~\ref{optoappendix}, the power requirements are optimized by matching the laser frequency to the fundamental frequency of the optical cavity (zero detuning).

For the cavity optomechanics system given above, the total measurement noise for the sensor operating at SQL at frequency $\omega_{\phi}$ is given by the simple formula (see equation \eqref{SQLFF})
\be
\label{SQLFF-main}
S_{FF}^{M,SQL}(\omega_{\phi}) = 2  m_s \sqrt{(\omega_{\phi}^2 - \omega_s^2)^2 + \gamma^2 \omega_s^2}.
\ee
Here $\gamma$ is the same mechanical damping rate appearing in \eqref{noise-thermal}. Note that this equation is true for force signals only at frequency $\omega_{\phi}$. Achieving the SQL is a frequency-dependent statement: one minimizes the measurement added noise at only the specific frequency $\omega_{\phi}$. The noise power at other frequencies $\omega \neq \omega_{\phi}$ is above the SQL. Thus if one wants to achieve the SQL-limited sensitivity over a range of possible signal frequencies, one needs to tune the laser power frequency-by-frequency while scanning over the range. See section \ref{bounds} for a more detailed discussion of search strategies along these lines.

We emphasize that \eqref{SQLFF-main} is true for arbitary signal frequency $\omega_{\phi}$. In particular, this equation holds for signals off resonance with the sensor $\omega_{\phi} \neq \omega_s$. It is clear from this equation that a better force sensitivity would come from not only achieving the SQL but also tuning the sensor to be on-resonance $\omega_{\phi} = \omega_s$, in which case we have
\be
\label{SQLFF-reso}
S_{FF}^{M,SQL,res}(\omega_{\phi}) = 2  m_s \gamma \omega_{\phi}.
\ee
Although tuning the laser power to achieve the SQL over a range of frequencies is in principle straightforward, tuning the actual mechanical resonance frequency is more challenging.\footnote{The ADMX \cite{du2018search} and HAYSTAC \cite{zhong2018results} experiments, for example, have implemented a search strategy precisely along these lines. There the resonance frequency of a microwave cavity sensor is tuned to resonantly search for axions over a range of frequencies.} One potentially promising avenue would be to use dynamical stiffening techniques, which allow one to increase the effective mechanical frequency by driving the system with laser backaction \cite{sheard2004observation}.

Environmental thermal noise is an irreducible floor for any measurement. Although one can lower the temperature, gas pressure, and so forth, any experiment will always have a thermal noise floor set by \eqref{noise-thermal}. This is in contrast to measurement-added noise, which can be reduced significantly \emph{below} the SQL by the use of sophisticated techniques like the use of squeezed readout light \cite{caves1981quantum,aasi2013enhanced,purdy2013strong,clark2017sideband} or quantum backaction-evasion \cite{braginsky1980quantum,BRAGINSKY1990251,clerk2008back,hertzberg2010back,khosla2017quantum,danilishin2018new}. In this paper, we will use the SQL level of measurement noise as a benchmark, but we emphasize that further detection reach is achievable using available post-SQL measurement techniques.

\subsection{Detection with an array}
\label{arrays}

All of our considerations above were for a single force sensor. Suppose now that we have an array of $N_{\rm det}$ such sensors. The DM signal is now a correlated force acting across the entire array. This provides two key advantages: increasing the signal-to-noise above any uncorrelated noises, as well as offering some routes to elimination of backgrounds and some correlated noises.

Consider first noise sources which act separately on each sensor. For example, we assume that both thermal noise and measurement-added noise, discussed above, have this property. The achievable force sensitivity with an array of $N_{\rm det}$ sensors is then simply given by $\sqrt{N_{\rm det}}$ times the single-sensor sensitivity, because the noises can be averaged out across the array. 

The $\sqrt{N_{\rm det}}$ enhancement comes from assuming that we read out each sensor individually. Further improvement can be made using a coherent readout scheme, in which the same light probes multiple sensors at the same time (note however this will also induce correlated measurement-added noise). In such a scheme, the signal adds at the level of the amplitude, which is then squared to produce the probability we actually read out. In the ultimate case that we use a single light probe to read out the entire array, one can achieve a full $N_{\rm det}$ enhancement--the so-called Heisenberg limit in quantum metrology \cite{giovannetti2011advances}.

Use of an array also serves an extremely important purpose beyond increasing the sensitivity: elimination of backgrounds and technical noise. In particular, as emphasized above and in \cite{graham2016dark}, the kinds of signal forces here couple to different materials differently (e.g., the $B-L$ force couples to neutron number). This is much like an equivalence principle (EP) violating force. If we construct an array consisting of sensors built from two different materials, we can then make differential measurements between the two types in order to remove backgrounds that act identically on all the sensors. Atom interferometers using two atomic species or torsion balance experiments looking for EP violation use this technique to remove seismic noise, which becomes extremely important at low frequencies. Other correlated noise sources can also be suppressed by doing differential measurements of this type. In the following section, we assume that these correlated noise sources have been sufficiently controlled so that thermal and measurement-added noise are dominant.

\section{Detection reach and search strategy}
\label{bounds}

\begin{figure}[t]
    \includegraphics[width=0.49\linewidth]{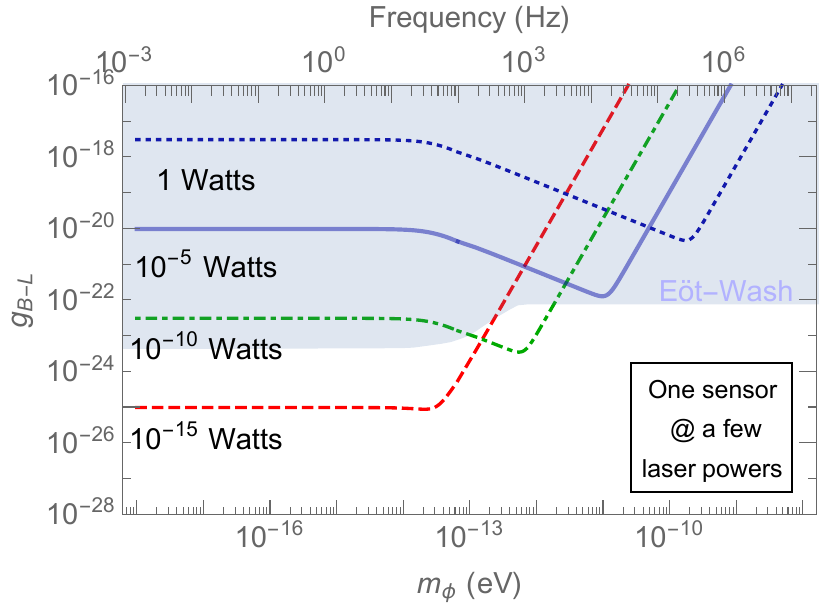} 
    \includegraphics[width=0.49\linewidth]{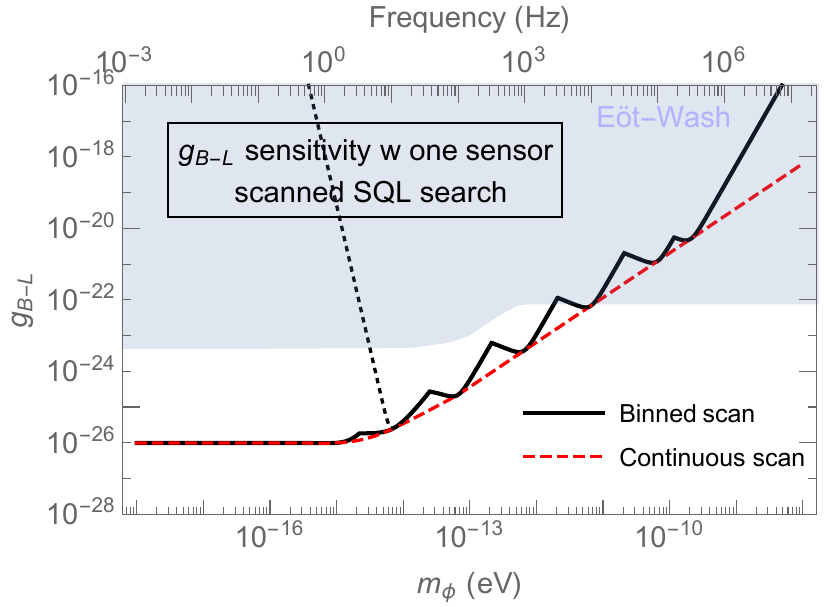}
    \includegraphics[width=0.49\linewidth]{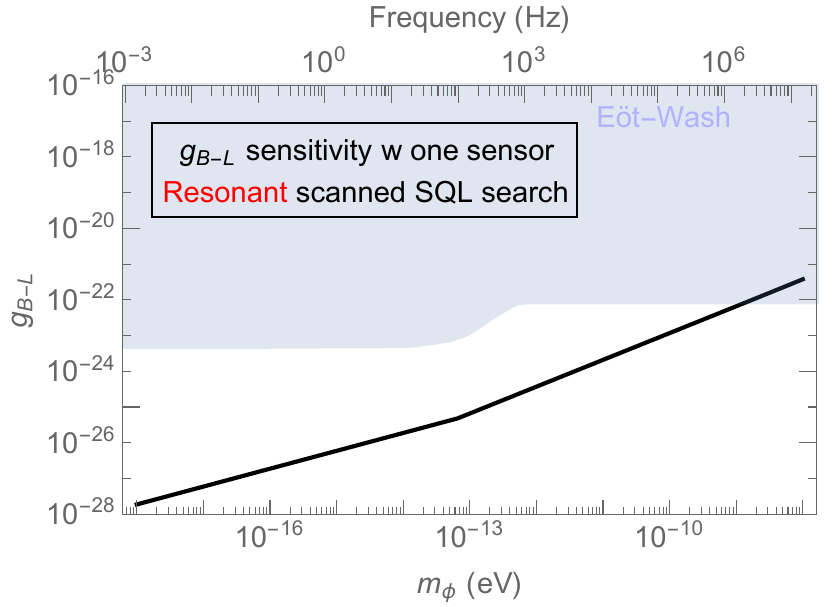}
    \caption{Sensitivity coverage on the dark matter coupling for various search strategies, using same sensor parameters as in figure \ref{figure-noisecontribs}, using the vector $B-L$ model for illustration. We assume a reference mirror made of iron and moveable mirror made of silicon for concreteness (again based on the setup in \cite{matsumoto2019demonstration}), leading to a differential acceleration coefficient $\Delta \approx 0.03$. Top left: sensitivity curves for the simple search strategy, with a single fixed laser power. Top right: idealized search strategy with sensor tuned to reach the SQL for each target frequency, with fixed mechanical frequency $\omega_s$. The jagged curve represents our binned search strategy described in the text. Bottom panel: fundamental limit on detection reach, scanning over resonance frequencies $\omega_{\phi} = \omega_s$. In all plots, we only allow for realistic laser powers $P_L \lesssim 1~{\rm W}$ as described in appendix \ref{optoappendix}.  Coupling strengths in the greyed-out region are already excluded by Eotvos torsion balance experiments~\cite{adelberger2009torsion,wagner2012torsion,graham2016dark}.}
    \label{figure-limits}
\end{figure}

In this section, we study the dark matter detection reach of mechanical sensor arrays. Given a specific sensor, environment, and measurement protocol, we have a noise PSD $S_{FF}(\omega)$. For simplicity, we will assume that the entirety of the DM sector consists solely of one ultralight field\footnote{This assumption can be relaxed by simply rescaling $F_0 = \sqrt{\rho_{\rm DM}}$ if the dark matter we are probing is a component of the full DM relic or we have local DM density fluctuations.} and take the force of the form in \eqref{forcegeneral}. We then obtain a bound on the coupling strength using our force sensitivity \eqref{sensitivity}. All told, with $N_{\rm det}$ sensors in the array, our fundamental DM detection reach (i.e. the smallest value of the DM-SM coupling we can detect) is set by
\be
\label{sensitivity-coupling-long}
g = \sqrt{\frac{S_{FF}(\omega_{\phi})}{N_{\rm det} N_g^2 F_0^2 T_{\rm tot}}}.
\ee
We assume incoherent readout of the individual sensors in the array; as discussed in section \ref{arrays}, coherent readout enhances the denominator by another power of $N_{\rm det}$.

In \eqref{sensitivity-coupling-long}, the time $T_{\rm tot}$ is an effective total intergration time. In general, we assume that our sensors can be operated for some integration time $T_{\rm int}$ set by technical constraints like laser stability, typically on the timescale of hours. There are then two basic DM regimes. Recall that the DM signal is essentially a coherent, monochromatic force on timescales $T_{\rm coh} \approx 10^6/\omega_{\phi}$ with $\omega_{\phi} = m_{\phi}$ set by the DM mass. For low frequency DM, we have $T_{\rm int} < T_{\rm coh}$, so a complete integration run will see a coherent signal. For higher-frequency DM candidates, $T_{\rm int} > T_{\rm coh}$, so the signal consists of $N_{\rm bins} = T_{\rm int}/T_{\rm coh}$ bins worth of independent signals. These bins can be summed in quadrature, and so we define the effective integration time
\be
T_{\rm tot} = \begin{cases} T_{\rm int} & T_{\rm int} < T_{\rm coh} \\ \sqrt{T_{\rm int} T_{\rm coh}} & T_{\rm int} > T_{\rm coh} \end{cases}.
\ee
We note that to scan for signals in bins in this fashion will require some use of template matching algorithms, similar to waveform matching used by LIGO. In particular, with sufficiently long integration times, templates will be required to take into account long-wavelength features like the rotation of the sensing setup with respect to the dark matter direction. While this can present computational challenges with long integration times, it also enables a mechanism for rejection of backgrounds which do not share the same temporal shape as the expected DM signal.

In the following, we consider three basic search strategies:

{\bf Single shot search}. The simplest ``strategy'' would be to simply assume a fixed laser power and integrate for $T_{\rm int}$. This produces a sensitivity curve which is best at some particular DM mass $m_{\phi}$, i.e. a particular signal frequency $\omega_{\phi}$, namely the frequency at which this laser power corresponds to the SQL. For DM masses away from this particular frequency, the resulting sensitivity curve is non-optimal. Some examples are given in figure \ref{figure-limits}(a). 

{\bf SQL scan at fixed mechanical frequency}. To do better, one should scan over dark matter masses by scanning over various values of the laser power, achieving the SQL frequency-by-frequency. Using \eqref{SQLFF-main} in \eqref{sensitivity-coupling-long} and assuming that the dominant thermal noise is from phononic coupling, this corresponds to a sensitivity curve
\be
\label{scan-sql}
g^2 =\frac{\gamma m_s k T + 2  m_s \sqrt{(\omega_s^2 - \omega_{\phi}^2)^2 + \gamma^2 \omega_{s}^2}}{N_{\rm det} N_g^2 F_0^2 T_{\rm tot}}.
\ee

In figure \ref{figure-limits}(b), we plot the ideal case in which one achieves the SQL for every frequency; since this is an infinite number of frequencies and we need to integrate for $T_{\rm int}$ at each, this is only a fundamental limit. A more realistic strategy is also plotted in \ref{figure-limits}(b), in which we bin the DM masses by order of magnitude, and tune the laser to a power optimized once for each bin. For the large span of available light DM parameter space below O($10^{-8}$ eV), and an integration time of order one hour per bin, the full spectrum could be scanned in roughly one day. In each case, we assume a maximum achievable laser power of around $1~{\rm W}$. This means that at sufficiently high DM mass, we are operating at a noise level worse than the SQL (see figure \ref{figure-laserreqs}).

{\bf Resonant SQL scan}. Finally, we consider the ultimate limit of resonant sensors operating at the SQL: a scan in which we vary the mechanical frequency of the sensor to resonance with the DM frequency $\omega_s = \omega_{\phi}$. From \eqref{scan-sql} this yields a sensitivity curve
\be
\label{scan-reso}
g^2 =\frac{\gamma m_s (k T + 2  \omega_{\phi})}{N_{\rm det} N_g^2 F_0^2 T_{\rm tot}}.
\ee
As discussed in section \ref{measurementnoise}, this could potentially be achieved (at least over some orders of magnitude in frequency) using dynamical stiffening of a fixed sensor.  

Equation \eqref{scan-reso} can be used to gain some basic intuition about the scaling of the DM detection problem. Note that $N_g \sim m_s$, so our smallest detectable DM coupling scales like $g \sim 1/\sqrt{N_{\rm det} m_s} = 1/\sqrt{m_{total}}$. Thus the fastest way to win is simply to build a more massive sensor! As discussed in section \ref{arrays}, the array approach can do better than simple scaling in the total mass if one has coherent readout of the array, in which case we can achieve a scaling $g \sim 1/\sqrt{N_{\rm det}^2 m_s}$. We also see that the overall sensitivity $g \sim \sqrt{\gamma}$, so lower damping rates (i.e. higher $Q$-factor resonators) can significantly increase our sensitivity. Finally, we can observe a simple crossover behavior: for low frequency signals $k T >  \omega_{\phi}$ the thermal noise dominates while for high frequencies $k T <  \omega_{\phi}$, the SQL measurement-added noise dominates. This in particular suggests that backaction-evasion or some other kind of post-SQL strategy will be most beneficial at frequencies $\omega_{\phi} \gtrsim k T \sim 1~{\rm GHz}$ (for dilution refrigeration temperatures $T \sim 10~{\rm mK}$).

\section{Conclusions and outlook}

Quantum sensing technology offers a highly promising route to searches for new physics beyond the standard model. Macroscopic sensors have already been demonstrated as gravitational wave detectors in LIGO, and development of these technologies should continue to push the precision frontier forward for years to come. 

In this paper, we have suggested a clear target for sensing using mechanical sensing devices: models of dark matter consisting of very light bosonic fields. If the dark sector contains such a field, it produces a nearly monochromatic force signal on a sensor, precisely the type of signal these sensors are optimized to detect. We have argued that sensors with already-demonstrated sensitivities have non-trivial detection reach in this parameter space. With an array of sensors, one can achieve a rapidly scaling sensitivity to these dark matter candidates. Since these signals also act different on different materials, the use of at least two material species of sensors can be used to make differential measurements, eliminating many systematic and background errors \cite{graham2016dark}. In Fig. \ref{figure-limits}, we see clearly that mechanical sensors are poised to make a contribution in the kHz-MHz regime, which would provide a complement to atom interferometer and torsion balance approaches operating at sub-kHz frequencies.

Beyond these ultra-light models, quantum-limited force sensing devices should be able to search for many other types of new physics. Since the sensors are macroscopic in size, they should have exquisite sensitivity to any potential signal whose strength increases with the mass of the sensor. Different realizations of the optomechanical force sensing paradigm could be used in complementary ways; for example, magnetic sensors could be used to search for dipole-coupled forces. Ultimately, a large array of these sensors could be able to search for heavy dark matter candidates purely through the gravitational interaction \cite{Carney:2019pza}. This paper presents a first example of a detection target reachable with currently demonstrated mechanical sensing technology, and we hope it encourages further exploration of the potential uses of quantum sensing for physics beyond the standard model.

\emph{Acknowledgements.}--We would like to thank Peter Graham, Roni Harnik, Gordan Krnjaic, Konrad Lehnert, Nobuyuki Matsumoto, David Moore, Cindy Regal, and Fengwei Yang for discussions. Special thanks to Jon Pratt for taking and providing us with some benchmark low-frequency seismic noise data. ZL, AH, and YZ would like to thank PITT-PACC, KITP and MIAPP for support from their programs and providing the environment for collaboration during various stages of this work. AH and ZL are supported in part by the NSF under Grant No. PHY1620074 and by the Maryland Center for Fundamental Physics. Y.Z. is supported by U.S. Department of Energy under Award Number DESC0009959. The Joint Quantum Institute is supported by funding from the NSF.

\appendix

\section{DM signal power spectral density}
\label{app-DMPSD}

Previously we made a simple estimate for the DM signal as a purely monochromatic force. Here, we give a more detailed discussion on the DM signal power spectral density if the frequency resolution is better than the intrinsic width caused by DM viral velocity.

The dark matter signal can be simulated by linearly adding up many freely propagating dark matter wavefunctions, where each dark matter wavefunction is described by a planewave.  For simplicity, we will describe the case of vector dark matter where
\begin{equation}
    \vec A (t,\vec x)= \sum_i \vec A_i \sin[\omega_i t-\vec k_i\cdot \vec x_i+\phi_i] .
\end{equation}
The polarization vector ($\vec A_i$) and the propagation vector ($\vec k_i$) are independent to each other. We take both vectors to be isotropic in the galaxy frame. In the non-relativistic limit, $\vec k = m \vec v$ and the magnitude of $\vec v_i$ follows Boltzmann distribution, i.e. $f(\vec v)\sim e^{-|v|^2/v_0^2}\Theta(v_{\rm escape}-|v|)$. Here $v_0\simeq 230~\rm km/s$ and $v_{\rm escape}\simeq 325~\rm km/s$. The frequency $\omega_i$ is determined by $|k_i|$ through the dispersion relation. The magnitude of $\vec A_i$, $A_n = |\vec A_i|$, is the same for all particles and it can be calculated by local DM energy density, i.e. around $0.4 ~\rm GeV/\rm cm^3$. Furthermore, $\phi_i$ is a random phase associated to each particle, which runs from 0 to $2\pi$ with uniform probability distribution (see \cite{Guo:2019ker} for details).   


The overall force on a sensor induced by vector dark matter can be written as
\begin{equation}
    \vec f \simeq N_{B-L} g_{B-L} \vec E \simeq  N_{B-L} g_{B-L} \sum_i \partial_t \vec A_i 
\end{equation}
In the lab frame, the force can be decomposed in Cartesian coordinate. However, this frame is moving with the Earth respect to the galaxy frame. This includes both the motion around the galactic center ($v_E\sim$230 km/s)
as well as the motion induced by the Earth's rotation, characterized by $\omega_E$. 
The effect induced by $v_E$ can be easily taken into account by shifting the velocity distribution by a constant vector. Meanwhile, the Earth rotation gives more interesting features to the signal.

Let us introduce the geodetic frame where the origin is located at the center of the Earth and the z-axis points to the north pole.  When comparing experiments done in different geological locations and when taking into account astrophysical effects, the geodetic frame is more natural.  This frame rotates with the Earth, thus one needs a constant rotation matrix, $R_{ab}$, to translate quantities in the geodetic frame to the lab frame. In the geodetic frame, the force induced by each plane wave can be written as
\begin{eqnarray}
       f_{G,x,i} &=& N_{B-L} g_{B-L} \omega_i A_n \sin(\theta_{A,i})\cos(\phi_{A,i}-\omega_E t) \nonumber\\
&&\times\sin(\omega_i t -\vec k_i\cdot \vec x_s+\phi_i) \nonumber\\
    &\simeq& N_{B-L} g_{B-L} \omega_i A_n \sin(\theta_{A,i})\cos(\phi_{A,i}-\omega_E t) \nonumber\\
&& \times \sin(\omega_i t +\phi_i) \nonumber\\
    f_{G,y,i} &\simeq& N_{B-L} g_{B-L} \omega_i A_n \sin(\theta_{A,i})\sin(\phi_{A,i}-\omega_E t) \nonumber\\
&& \times \sin(\omega_i t +\phi_i) \nonumber\\
    f_{G,z,i} &\simeq& N_{B-L} g_{B-L} \omega_i A_n \cos(\theta_{A,i})\sin(\omega_i t +\phi_i)  
\end{eqnarray}
Here $\vec x_s$ is the position vector of the sensor in the geodetic frame. In the parameter space that we are interested in, the de Broglie wavelength can sometimes be larger than the size of the Earth. For example, when the signal frequency is about 100 Hz, the de Broglie wavelength is $O(10^9)$m, which is much larger than the radius of the Earth. In addition, when we increase signal frequency, the Earth rotation effect becomes less important. In that limit, $\vec k_i\cdot \vec x_s$ can be treated as a constant and absorbed into the random phase.  Working in this limit, we can safely drop the $\vec k_i\cdot \vec x_s$ term. Furthermore, $(\theta_A, \phi_A)$ are the polar angles of $\vec A_i$ in the geodetic frame at $t=0$.  It is interesting to note that the Earth's rotation can split a monochromatic signal into three frequencies, with frequency spacing as $\omega_E$.

Translating into the lab frame with the assumption that the measurement is along the x-axis in the lab frame, we have 
\begin{eqnarray}
       f_{s,x} =\sum_i (R_{11} f_{G,x,i}+R_{21} f_{G,y,i}+R_{31} f_{G,z,i}).
\end{eqnarray}
The sum over $i$ randomly averages the DM field variables, leading to an overall amplitude $f \propto \sqrt{N_{waves}} \propto \sqrt{\rho_{\rm DM}}$, as in our general parametrization \eqref{forcegeneral}. In figure \ref{fig:signal} we show the behavior of the signal power spectrum as a function of frequency. Here we choose vector dark matter mass so that the oscillation frequency is around 1kHz and we linearly add one million vector dark matter wavefunctions.\footnote{We checked that increasing the number of vector dark matter does not change our result qualitatively.} We assume the total observation time is 200 hours, which gives frequency resolution as $\sim 1.4 \mu$Hz. The spiky feature in this power spectral density (PSD) is caused by the incoherent superposition of the particles in one frequency bin. By combining 125 bins togehter, we obtain the averaged PSD, shown as blue dots in figure \ref{fig:signal}.

\section{Continuous force sensing with optomechanics}

\label{optoappendix}

Here we review some basics about continuous force sensing in optomechanical and electromechanical systems. We will use a single-sided optical cavity coupled to a mechanical object as a basic tool; many other systems obey the same basic equations. We begin with a continuous position sensing setup and then discuss velocity measurements. Our discussion will be reasonably self-contained, but we refer the reader to eg. the review \cite{clerk2010introduction} for details, especially on the equations of motion for the cavity.

\subsection{Single-sided optomechanical cavity}

Consider a high-reflectivity mirror suspended as a mechanical resonator of mechanical frequency $\omega_s$ and mass $m_s$ forming a moveable end of an optical cavity with a single port. We write the total Hamiltonian
\be
H = H_{sys} + H_{bath}.
\ee
The system term here consists of a mechanical resonator (the sensor) and the fundamental electromechanical mode of the cavity. There are two baths, one each for the mechanics and cavity. The mechanical bath can be thought of as phonons in the support structure of the resonator mass, while the cavity bath consists of photonic modes in a transmission line connected to the input/output port. In particular, the cavity bath will also serve to drive the system and read out measurements. 

The Hamiltonian of the mechanical oscillator and cavity mode are
\be
H_{sys} = \omega a^{\dagger} a +  \omega_s b^{\dagger} b.
\ee
The first term is the cavity mode and the second is the mechanical oscillator. We will often use the cavity amplitude and phase quadrature variables, defined by 
\be
\label{quadratures}
X = \frac{a + a^{\dagger}}{\sqrt{2}}, \ \ \ Y = -i\frac{a - a^{\dagger}}{\sqrt{2}}
\ee
respectively. The mechanics and light are coupled optomechanically: the frequency $\omega$ of the cavity depends on the oscillator position $\omega = \omega(x)$. Taylor expanding this dependence to lowest order gives the kinetic term with $\omega(0) =: \omega_c$, the cavity frequency at $x=0$. The next order gives the coupling
\be
\label{Hint}
H_{int} =  G_0 \frac{x}{x_0} a^{\dagger} a, \ \ \ G_0 = -\frac{x_{0} \omega_c}{L}.
\ee
where $L$ is the equilibrium length of the cavity and $x_{0} = \sqrt{1/2 m \omega_s}$ is the zero point length scale of the mechanical oscillator. Note this is normalized so that $G_0$ has units of a frequency.\footnote{We will use uppercase $G$ for the optomechanical coupling to avoid confusion with the lowercase $g$ used for the DM-sensor coupling in \eqref{forcegeneral}.} This coupling is key to optomechanics, allowing us to prepare and read out the mechanical state using the cavity photons. 

Driving the system with an external laser amounts to displacing the cavity mode operators
\be
a \to e^{i \omega_L t} (\alpha + a),
\ee
where $\omega_L$ is the laser frequency, which we scale out of the cavity operators (i.e. we work in the frame co-rotating with the laser). The classical laser drive strength $\alpha \sim \sqrt{P_L}$ leads to an enhanced optomechanical coupling strength $G$:
\be
\label{couplingconstant}
G = G_0 \sqrt{\oL{n}}, \ \ \ \oL{n} = \frac{P_L}{\omega_L \kappa},
\ee
where $G_0$ is the single-photon optomechanical coupling given in \eqref{Hint} and $\oL{n}$ is the average occupancy of the cavity mode in terms of the laser power $P_L$ and cavity loss $\kappa$. Taking the drive $\alpha$ to be real and linearizing around this large background field, we obtain the optomechanical coupling
\be
H_{OM} =  G \frac{x}{x_0} X.
\ee
The drive changes the equilibrium position and frequency of the oscillator to linear order; we have redefined $\omega_c, \omega_s$ to include this effect.

Finally, we need to discuss the effects of the baths. Both the mechanical and cavity oscillators are (separately) coupled to large baths of bosonic modes
\begin{align}
\begin{split}
H_{bath} & = \sum_p \omega_p A_p^{\dagger} A_p + \nu_p B_p^{\dagger} B_p \\
& - x \sum_p g_p x_p - i \sum_p f_p a^{\dagger} A_p - f_p^* a A_p^{\dagger}.
\end{split}
\end{align}
Note that the cavity-bath and sensor-bath couplings are different: in particular, the sensor bath couples only to the mechanical position $x$. The mechanical bath is unmonitored, leading to simple damping of the mechanics. We now make the usual Markovian approximation for the bath and assume that the bath-system couplings $f_p \equiv f, g_p \equiv g$ are approximately constant. Tracing out the baths then leads to the Heisenberg-Langevin equations of motion for the cavity and mechanics
 \cite{clerk2010introduction}
\begin{align}
\begin{split}
\label{EOM}
\dot{X} & = -\Delta Y - \frac{\kappa}{2} X + X_{in} \\
\dot{Y} & = \Delta Y - \frac{\kappa}{2} Y + Y_{in} + \frac{G x}{x_0} \\
\dot{x} & = \frac{p}{m_s} \\
\dot{p} & = - m_s \omega_s^2 x - \gamma p + F_{in} + \frac{G X}{x_0}.
\end{split}
\end{align}
Here $\Delta = \omega_L - \omega_c$ is the cavity detuning from the laser drive at frequency $\omega_L$ and $\kappa, \gamma$ are the damping rates of the cavity and mechanics respectively. In what follows we take the detuning $\Delta = 0$. 

The input fields $X_{in}, Y_{in}, F_{in}$ are operators which represent the input noise sources. They are constructed as sums over the individual bath modes, eg. 
\be
\label{Ain}
A_{in}(t) = \frac{1}{\sqrt{2 \pi \rho_A}} \sum_p e^{-i \omega_p(t-t_0)} A_{p}(t_0)
\ee
represents the cavity input annihilation operator, where $\rho_A$ is the density of states of the transmission line and $A_{p}(t_0)$ is the intial condition of the mode. From this we can construct $X_{in}, Y_{in}$ in analogy with \eqref{quadratures}. Physically, the $X_{in},Y_{in}$ represent quantum vacuum noise around the laser drive and $F_{in}$ represents a thermal bath coupling to the mechanics. The optical noises satisfy
\be
\label{XYnoise}
[X_{in}(t), X_{in}(t') ] = [Y_{in}(t), Y_{in}(t') ] = \kappa \delta(t-t'),
\ee
representing vacuum white noise from a 1D QFT; note that they have dimensions of $1/\text{time}$. The input force noise satisfies
\be
\label{Fnoise}
[F_{in}(t), F_{in}(t') ] = \alpha \delta(t - t'), \ \ \ \alpha = 4 \gamma m_s k T,
\ee
with $T$ the temperature of the mechanical bath.\footnote{The appearance of the same damping coefficient between the $X$ and $X_{in}$ operators (similarly $Y,F$) is a consequence of the fluctuation-dissipation theorem. It is a reflection of the assumption that the cavity and mechanics are in equilibrium with their respective baths.} Thus $F_{in}$ has dimensions of $\text{force} = \text{momentum}/\text{time}$.

\subsection{Inferring force from position measurement}

\begin{figure}[t]
\includegraphics[width=0.46\linewidth]{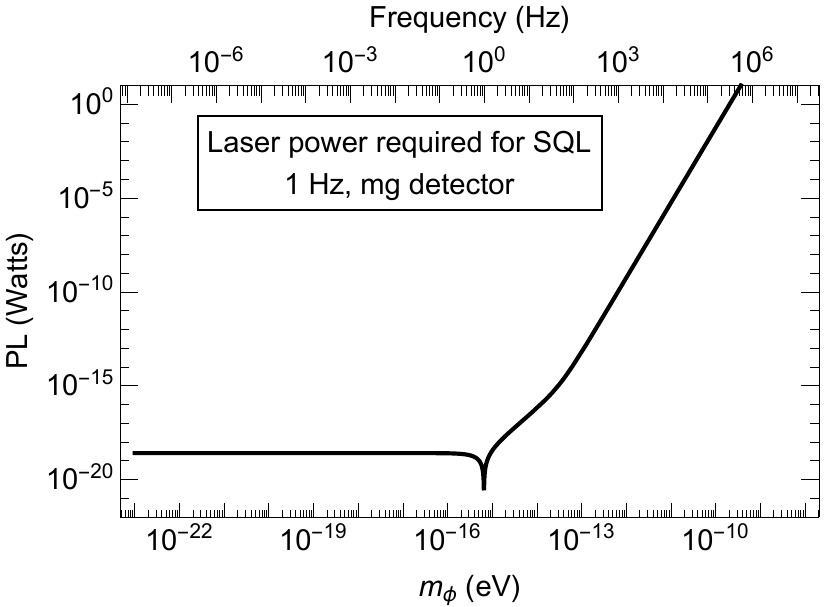} 
\includegraphics[width=0.46\linewidth]{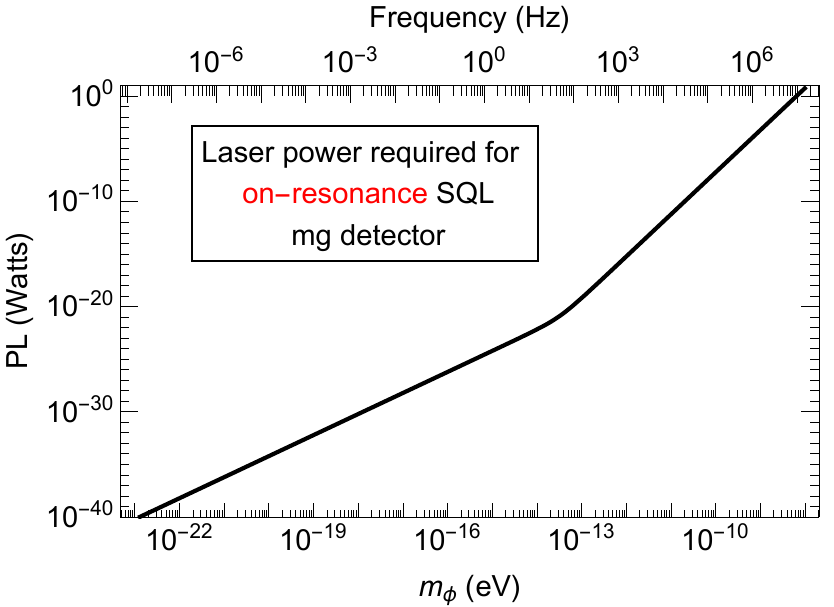}
\caption{Laser power required to achieve the SQL. Top: fixed mechanical frequency $\omega_s$. Bottom: variable mechanical frequency $\omega_s$, with SQL achieved on-resonance $\omega_{\phi} =\omega_s$. Same sensor parameters as in figure \ref{figure-noisecontribs}, and assuming a cavity length $L = 0.1~{\rm m}$. We see that at sufficiently high signal frequency, the SQL requires prohibitively large laser power $P_L \gtrsim 1~{\rm W}$. In our sensitivity curves, we only allow for laser powers below $1~{\rm W}$; at higher frequencies we are therefore working at noise levels above the SQL.}
\label{figure-laserreqs}
\end{figure}

The standard paradigm for force sensing is continuously monitor some resonator variable like $x(t)$ and try to infer a force acting on the device. Note that $x(t)$ is imprinted onto the light $Y$ quadrature through the optomechanical coupling; what one actually does is to make an interferometric measurement of this quadrature.

To imprint the mechanical information onto the light, consider an incoming photon which scatters off the mirror and then comes back out of the cavity. The light picks up a phase shift proportional to $x(t)$ where $t$ is the time of scattering. The full scattering process involves the photon being shot into the cavity, reflected, shot back out of the cavity, and finally measured. Thus we have essentially an $S$-matrix style description; the way people discuss this in quantum optics is through the so-called input-output relations
\be
\label{in-out}
X_{out} = X_{in} + \kappa X, \ \ \ Y_{out} = Y_{in} + \kappa Y.
\ee
This is the same way one formulates scattering through the LSZ formula. The output fields are defined in analogy with the input fields like \eqref{Ain}, but with respect to the late-time values of the mode operators. Again, see \cite{clerk2010introduction} for details.

The noise in some observable $\Oi(t)$ has power spectral density (PSD) determined by
\be
\label{npsd-definition}
\braket{ \Oi(\omega) \Oi(\omega') } = S_{\Oi \Oi}(\omega) \delta(\omega+\omega').
\ee
The delta function arises by assumption of a stationary noise source, which is accurate in our case. Let $[\Oi]$ be the dimension of $\Oi(t)$, so $\Oi(\omega)$ has units of $[\Oi] \times \text{time}$, and thus $S_{FF}(\omega)$ has units of $[\Oi]^2/\text{frequency}$. The square root is what gives the usual measure of sensitivity like LIGO's strain per root Hertz or an accelerometer's g per root Hertz. The interpretation is that if one wants to measure $\Oi$ to some precision $\Delta \Oi$, you need to integrate for a time $T_{\rm int}$ given by $\Delta \Oi = \sqrt{S_{\Oi\Oi}/T_{\rm int}}$. The more accurate (small $\Delta \Oi$), the larger $T_{\rm int}$ is needed.

Thus, we need two key pieces of information to determine our sensitivity: a choice of operator and a calculation of the noise PSD. As described above, in our case the observable we actually measure is the output phase quadrature $Y_{out}$. Let us calculate its noise. The goal is to use the equations of motion \eqref{EOM} and input-output relations \eqref{in-out} to compute the noise PSD \eqref{npsd-definition} with $\Oi = Y_{out}$. The answer should be expressed in terms of the vacuum input noises $Y_{in}, X_{in}, F_{in}$ only. 

The equations of motion \eqref{EOM} are linear so this can be performed with a page of algebra. Working in the frequency domain, we have
\begin{align}
\begin{split}
\label{XY1}
X(\omega) & = \chi_c(\omega) X_{in}(\omega) \\
Y(\omega) & = \chi_c(\omega) \left[ Y_{in}(\omega) + \frac{G x(\omega)}{x_0} \right]
\end{split}
\end{align}
where the cavity susceptibility is 
\be
\label{susc-cavity}
\chi_c(\omega) = \frac{1}{i \omega- \kappa/2}.
\ee
The oscillator position spectrum is
\be
\label{x1}
x(\omega) = \chi_{m}(\omega) \left( F_{in}(\omega) + \frac{ G X(\omega)}{x_0} \right)
\ee
with the mechanical susceptibility 
\be
\label{susc-mechanics}
\chi_m(\omega) = \frac{1}{m_s(\omega_s^2-\omega^2 - i \gamma \omega_s)}.
\ee
Plugging \eqref{x1} into \eqref{XY1} and using the input-output relations \eqref{in-out}, we obtain the output phase quadrature
\begin{align}
Y_{out} = e^{i \phi_c} Y_{in} + \frac{ G^2 \kappa}{x_0^2} \chi_c^2 \chi_m X_{in} + \frac{G \kappa}{x_0} \chi_c \chi_m F_{in},
\end{align}
where $e^{i \phi_c} = 1 + \kappa \chi_c$. From this equation one can easily work out the $Y$ noise PSD by inserting this into \eqref{npsd-definition} and using the vacuum correlation function of the input optical noises \eqref{XYnoise} and thermal correlation function of the input force noise \eqref{Fnoise}.

In force sensing, what one really does in practice is filter $Y$ with the appropriate filtering function to reference all of this to force. In this case that means we define our force estimator
\be
F_E(\omega) = Y_{out}(\omega) \frac{x_0}{G \kappa \chi_c(\omega) \chi_m(\omega)}.
\ee
Defining the noise PSD $S_{FF}(\omega)$ of our force readout using \eqref{npsd-definition} with $\Oi = F_E$, we obtain
\be
S_{FF}(\omega) = S_{FF}^{SN}(\omega) + S_{FF}^{BA}(\omega) + S_{FF}^{T}(\omega)
\ee
with
\begin{align}
\begin{split}
S_{FF}^{SN} & = \frac{x_0^2}{G^2 \kappa}\frac{1}{|\chi_c \chi_m |^2} \\
S_{FF}^{BA} & = \frac{G^2 \kappa}{x_0^2} \left| \chi_c \right|^2 \\
S_{FF}^{T} & = 4 \gamma m_s k T.
\end{split}
\end{align}
Here we assumed that $\braket{ [X_{in},Y_{in}] }_{vac} = 0$, i.e. there is no correlation between the two quadratures. If we used squeezed input light (as LIGO has begun to do \cite{aasi2013enhanced}), this correlator does not vanish and can actually contribute with a minus sign, i.e. lower the total noise. In what follows we will assume zero squeezing. 

The measurement-added noise $S^{M} = S^{SN} + S^{BA}$ consists of shot noise (fluctuations in the input phase $Y_{in}$) and backaction (fluctuations in the input amplitude $X_{in}$). Note that the shot noise (referenced to a force measurement!) goes like $1/P_L$ while backaction goes like $P$ with $P_L$ the input laser power. For a fixed frequency, one can therefore minimize the sum of the shot noise and backaction terms with some optimal laser power $P_{SQL}$ to achieve the so-called standard quantum limit (SQL). Note that this can only be done at a single frequency.

Let $\omega_{\phi}$ be a fiducial signal frequency at which we want to achieve the SQL. The SQL occurs at a coupling $G_* = G_*(\omega_{\phi})$ determined by $dS^{M}(\omega_{\phi})/dG^2 = 0$, namely
\be
G_*^2(\omega_{\phi}) = \frac{x_0^2}{\kappa} \frac{1}{|\chi^2_c(\omega_{\phi}) \chi_m(\omega_{\phi})|}.
\ee
This determines the requisite laser power $P_L$ through \eqref{couplingconstant}:
\be
P_L^{SQL}(\omega_{\phi}) = \frac{L^2}{\omega_c} \frac{1}{|\chi_c^2(\omega_{\phi}) \chi_m(\omega_{\phi})|}
\ee
assuming exactly zero detuning $\omega_c = \omega_L$. See figure \ref{figure-laserreqs} for an example of the required laser power as a function of target frequency $\omega_{\phi}$. At this value of the coupling, the measurement-added part of the noise PSD becomes
\begin{align}
S_{FF}^{M,SQL}(\omega | \omega_{\phi}) =  \left| \frac{\chi^2_c(\omega_{\phi}) \chi_m(\omega_{\phi})}{\chi^2_c(\omega) \chi^2_m(\omega)} \right| +  \left| \frac{\chi^2_c(\omega)}{\chi^2_c(\omega_{\phi}) \chi_m(\omega_{\phi})} \right|.
\end{align}
Note in particular that at the SQL frequency $\omega = \omega_{\phi}$ we have the simple result
\be
\label{SQLFF}
S_{FF}^{M,SQL}(\omega = \omega_{\phi}) = 2  m_s \sqrt{(\omega_{\phi}^2 - \omega_s^2)^2 + \gamma^2 \omega_s^2}.
\ee
This equation demonstrates clearly the ideal case: resonant detection, with the signal and mechanical frequencies matching $\omega_{\phi} = \omega_s$.

In the above, we have given our sensitivity curves based on the fundamental limits set purely by thermal and measurement-added noise. At low frequencies (below around $1~{\rm Hz}$), seismic noise starts to dominate over these two sources. As discussed in the main text, seismic noise can in principle be subtracted by using sensors with different material compositions and making differential measurements, which can distinguish between the seismic noise.and equivalence-principle violating dark matter signal. To get some quantitative sense of the magnitude of seismic noise at low frequencies, we show a typical $1/f$ spectrum normalized to data taken by Jon Pratt at NIST, Gaithersburg in the plots in Figs. \ref{figure-noisecontribs}, \ref{figure-limits}.

\bibliographystyle{utphys-dan} 
\bibliography{alp-det-refs}

\providecommand{\href}[2]{#2}\begingroup\raggedright\begin{thebibliography}{10}

\bibitem{Sofue:2000jx}
Y.~Sofue and V.~Rubin, ``{Rotation curves of spiral galaxies},''
  \href{http://dx.doi.org/10.1146/annurev.astro.39.1.137}{{\em Ann. Rev.
  Astron. Astrophys.} {\bfseries 39} (2001) 137--174},
\href{http://arxiv.org/abs/astro-ph/0010594}{{\ttfamily arXiv:astro-ph/0010594
  [astro-ph]}}.

\bibitem{Massey:2010hh}
R.~Massey, T.~Kitching, and J.~Richard, ``{The dark matter of gravitational
  lensing},'' \href{http://dx.doi.org/10.1088/0034-4885/73/8/086901}{{\em Rept.
  Prog. Phys.} {\bfseries 73} (2010) 086901},
\href{http://arxiv.org/abs/1001.1739}{{\ttfamily arXiv:1001.1739
  [astro-ph.CO]}}.

\bibitem{Aghanim:2018eyx}
{\bfseries Planck} Collaboration, N.~Aghanim {\em et~al.}, ``{Planck 2018
  results. VI. Cosmological parameters},''
\href{http://arxiv.org/abs/1807.06209}{{\ttfamily arXiv:1807.06209
  [astro-ph.CO]}}.

\bibitem{Primack:2015kpa}
J.~R. Primack,
  \href{http://dx.doi.org/10.1017/9781316535783.008}{``{Cosmological Structure
  Formation},''} in {\em {The Philosophy of Cosmology}}, pp.~136--160.
\newblock 2015.
\newblock
\href{http://arxiv.org/abs/1505.02821}{{\ttfamily arXiv:1505.02821
  [astro-ph.GA]}}.
\newblock

\bibitem{Tanabashi:2018oca}
{\bfseries Particle Data Group} Collaboration, M.~Tanabashi {\em et~al.},
  ``{Review of Particle Physics},''
\href{http://dx.doi.org/10.1103/PhysRevD.98.030001}{{\em Phys. Rev.} {\bfseries
  D98} no.~3, (2018) 030001}.

\bibitem{Essig:2013lka}
R.~Essig {\em et~al.}, ``{Working Group Report: New Light Weakly Coupled
  Particles},'' in {\em {Proceedings, 2013 Community Summer Study on the Future
  of U.S. Particle Physics: Snowmass on the Mississippi (CSS2013): Minneapolis,
  MN, USA, July 29-August 6, 2013}}.
\newblock 2013.
\newblock \href{http://arxiv.org/abs/1311.0029}{{\ttfamily arXiv:1311.0029
  [hep-ph]}}.
\newblock
\url{http://www.slac.stanford.edu/econf/C1307292/docs/IntensityFrontier/NewLight-17.pdf}.
\newblock

\bibitem{Hui:2016ltb}
L.~Hui, J.~P. Ostriker, S.~Tremaine, and E.~Witten, ``{Ultralight scalars as
  cosmological dark matter},''
  \href{http://dx.doi.org/10.1103/PhysRevD.95.043541}{{\em Phys. Rev.}
  {\bfseries D95} no.~4, (2017) 043541},
\href{http://arxiv.org/abs/1610.08297}{{\ttfamily arXiv:1610.08297
  [astro-ph.CO]}}.

\bibitem{Irastorza:2018dyq}
I.~G. Irastorza and J.~Redondo, ``{New experimental approaches in the search
  for axion-like particles},''
  \href{http://dx.doi.org/10.1016/j.ppnp.2018.05.003}{{\em Prog. Part. Nucl.
  Phys.} {\bfseries 102} (2018) 89--159},
\href{http://arxiv.org/abs/1801.08127}{{\ttfamily arXiv:1801.08127 [hep-ph]}}.

\bibitem{adelberger2009torsion}
E.~Adelberger, J.~Gundlach, B.~Heckel, S.~Hoedl, and S.~Schlamminger, ``Torsion
  balance experiments: A low-energy frontier of particle physics,'' {\em
  Progress in Particle and Nuclear Physics} {\bfseries 62} no.~1, (2009)
  102--134.

\bibitem{wagner2012torsion}
T.~A. Wagner, S.~Schlamminger, J.~Gundlach, and E.~G. Adelberger,
  ``Torsion-balance tests of the weak equivalence principle,'' {\em Classical
  and Quantum Gravity} {\bfseries 29} no.~18, (2012) 184002.

\bibitem{graham2016dark}
P.~W. Graham, D.~E. Kaplan, J.~Mardon, S.~Rajendran, and W.~A. Terrano, ``Dark
  matter direct detection with accelerometers,'' {\em Physical Review D}
  {\bfseries 93} no.~7, (2016) 075029.

\bibitem{pierce2018searching}
A.~Pierce, K.~Riles, and Y.~Zhao, ``Searching for dark photon dark matter with
  gravitational-wave detectors,'' {\em Physical review letters} {\bfseries 121}
  no.~6, (2018) 061102.

\bibitem{Guo:2019ker}
H.-K. Guo, K.~Riles, F.-W. Yang, and Y.~Zhao, ``{Searching for Dark Photon Dark
  Matter in LIGO O1 Data},''
\href{http://arxiv.org/abs/1905.04316}{{\ttfamily arXiv:1905.04316 [hep-ph]}}.

\bibitem{Coleman:2018ozp}
{\bfseries MAGIS-100} Collaboration, J.~Coleman, ``{Matter-wave Atomic
  Gradiometer InterferometricSensor (MAGIS-100) at Fermilab},''
  \href{http://dx.doi.org/10.22323/1.340.0021}{{\em PoS} {\bfseries ICHEP2018}
  (2019) 021}, \href{http://arxiv.org/abs/1812.00482}{{\ttfamily
  arXiv:1812.00482 [physics.ins-det]}}.

\bibitem{du2018search}
N.~Du, N.~Force, R.~Khatiwada, E.~Lentz, R.~Ottens, L.~Rosenberg, G.~Rybka,
  G.~Carosi, N.~Woollett, D.~Bowring, {\em et~al.}, ``Search for invisible
  axion dark matter with the axion dark matter experiment,'' {\em Physical
  review letters} {\bfseries 120} no.~15, (2018) 151301.

\bibitem{zhong2018results}
L.~Zhong, S.~Al~Kenany, K.~Backes, B.~Brubaker, S.~Cahn, G.~Carosi,
  Y.~Gurevich, W.~Kindel, S.~Lamoreaux, K.~Lehnert, {\em et~al.}, ``Results
  from phase 1 of the haystac microwave cavity axion experiment,'' {\em
  Physical Review D} {\bfseries 97} no.~9, (2018) 092001.

\bibitem{TheLIGOScientific:2014jea}
{\bfseries LIGO Scientific} Collaboration, J.~Aasi {\em et~al.}, ``{Advanced
  LIGO},''
{\em Class. Quant. Grav.} {\bfseries 32} (2015) .

\bibitem{matsumoto20155}
N.~Matsumoto, K.~Komori, Y.~Michimura, G.~Hayase, Y.~Aso, and K.~Tsubono,
  ``5-mg suspended mirror driven by measurement-induced backaction,'' {\em
  Physical Review A} {\bfseries 92} no.~3, (2015) 033825.

\bibitem{matsumoto2019demonstration}
N.~Matsumoto, S.~B. Cata{\~n}o-Lopez, M.~Sugawara, S.~Suzuki, N.~Abe,
  K.~Komori, Y.~Michimura, Y.~Aso, and K.~Edamatsu, ``Demonstration of
  displacement sensing of a mg-scale pendulum for mm-and mg-scale gravity
  measurements,'' {\em Physical review letters} {\bfseries 122} no.~7, (2019)
  071101.

\bibitem{reinhardt2016ultralow}
C.~Reinhardt, T.~M{\"u}ller, A.~Bourassa, and J.~C. Sankey, ``Ultralow-noise
  sin trampoline resonators for sensing and optomechanics,'' {\em Physical
  Review X} {\bfseries 6} no.~2, (2016) 021001.

\bibitem{tsaturyan2017ultracoherent}
Y.~Tsaturyan, A.~Barg, E.~S. Polzik, and A.~Schliesser, ``Ultracoherent
  nanomechanical resonators via soft clamping and dissipation dilution,'' {\em
  Nature nanotechnology} {\bfseries 12} no.~8, (2017) 776.

\bibitem{yin2013optomechanics}
Z.-Q. Yin, A.~A. Geraci, and T.~Li, ``Optomechanics of levitated dielectric
  particles,'' {\em International Journal of Modern Physics B} {\bfseries 27}
  no.~26, (2013) 1330018.

\bibitem{chang2010cavity}
D.~E. Chang, C.~Regal, S.~Papp, D.~Wilson, J.~Ye, O.~Painter, H.~J. Kimble, and
  P.~Zoller, ``Cavity opto-mechanics using an optically levitated nanosphere,''
  {\em Proceedings of the National Academy of Sciences} {\bfseries 107} no.~3,
  (2010) 1005--1010.

\bibitem{hempston2017force}
D.~Hempston, J.~Vovrosh, M.~Toro{\v{s}}, G.~Winstone, M.~Rashid, and
  H.~Ulbricht, ``Force sensing with an optically levitated charged
  nanoparticle,'' {\em Applied Physics Letters} {\bfseries 111} no.~13, (2017)
  133111.

\bibitem{monteiro2017optical}
F.~Monteiro, S.~Ghosh, A.~G. Fine, and D.~C. Moore, ``Optical levitation of
  10-ng spheres with nano-g acceleration sensitivity,'' {\em Physical Review A}
  {\bfseries 96} no.~6, (2017) 063841.

\bibitem{chan1987superconducting1}
H.~Chan and H.~Paik, ``Superconducting gravity gradiometer for sensitive
  gravity measurements. i. theory,'' {\em Physical Review D} {\bfseries 35}
  no.~12, (1987) 3551.

\bibitem{chan1987superconducting2}
H.~Chan, M.~Moody, and H.~Paik, ``Superconducting gravity gradiometer for
  sensitive gravity measurements. ii. experiment,'' {\em Physical Review D}
  {\bfseries 35} no.~12, (1987) 3572.

\bibitem{prat2017ultrasensitive}
J.~Prat-Camps, C.~Teo, C.~Rusconi, W.~Wieczorek, and O.~Romero-Isart,
  ``Ultrasensitive inertial and force sensors with diamagnetically levitated
  magnets,'' {\em Physical Review Applied} {\bfseries 8} no.~3, (2017) 034002.

\bibitem{childress2017cavity}
L.~Childress, M.~Schmidt, A.~Kashkanova, C.~Brown, G.~Harris, A.~Aiello,
  F.~Marquardt, and J.~Harris, ``Cavity optomechanics in a levitated helium
  drop,'' {\em Physical Review A} {\bfseries 96} no.~6, (2017) 063842.

\bibitem{giovannetti2011advances}
V.~Giovannetti, S.~Lloyd, and L.~Maccone, ``Advances in quantum metrology,''
  {\em Nature photonics} {\bfseries 5} no.~4, (2011) 222.

\bibitem{Carney:2019pza}
D.~Carney, S.~Ghosh, G.~Krnjaic, and J.~M. Taylor, ``{Gravitational Direct
  Detection of Dark Matter},''
  \href{http://arxiv.org/abs/1903.00492}{{\ttfamily arXiv:1903.00492
  [hep-ph]}}.

\bibitem{Lisanti:2016jxe}
M.~Lisanti, \href{http://dx.doi.org/10.1142/9789813149441_0007}{``{Lectures on
  Dark Matter Physics},''} in {\em {Proceedings, Theoretical Advanced Study
  Institute in Elementary Particle Physics: New Frontiers in Fields and Strings
  (TASI 2015): Boulder, CO, USA, June 1-26, 2015}}, pp.~399--446.
\newblock 2017.
\newblock
\href{http://arxiv.org/abs/1603.03797}{{\ttfamily arXiv:1603.03797 [hep-ph]}}.
\newblock

\bibitem{Arias:2012az}
P.~Arias, D.~Cadamuro, M.~Goodsell, J.~Jaeckel, J.~Redondo, and A.~Ringwald,
  ``{WISPy Cold Dark Matter},''
  \href{http://dx.doi.org/10.1088/1475-7516/2012/06/013}{{\em JCAP} {\bfseries
  1206} (2012) 013},
\href{http://arxiv.org/abs/1201.5902}{{\ttfamily arXiv:1201.5902 [hep-ph]}}.

\bibitem{Armengaud:2017nkf}
E.~Armengaud, N.~Palanque-Delabrouille, C.~Yèche, D.~J.~E. Marsh, and J.~Baur,
  ``{Constraining the mass of light bosonic dark matter using SDSS
  Lyman-$\alpha$ forest},'' \href{http://dx.doi.org/10.1093/mnras/stx1870}{{\em
  Mon. Not. Roy. Astron. Soc.} {\bfseries 471} no.~4, (2017) 4606--4614},
\href{http://arxiv.org/abs/1703.09126}{{\ttfamily arXiv:1703.09126
  [astro-ph.CO]}}.

\bibitem{Marsh:2018zyw}
D.~J.~E. Marsh and J.~C. Niemeyer, ``{Strong Constraints on Fuzzy Dark Matter
  from Ultrafaint Dwarf Galaxy Eridanus II},''
\href{http://arxiv.org/abs/1810.08543}{{\ttfamily arXiv:1810.08543
  [astro-ph.CO]}}.

\bibitem{Bar:2019bqz}
N.~Bar, K.~Blum, J.~Eby, and R.~Sato, ``{Ultralight dark matter in disk
  galaxies},'' \href{http://dx.doi.org/10.1103/PhysRevD.99.103020}{{\em Phys.
  Rev.} {\bfseries D99} no.~10, (2019) 103020},
\href{http://arxiv.org/abs/1903.03402}{{\ttfamily arXiv:1903.03402
  [astro-ph.CO]}}.

\bibitem{Bar:2018acw}
N.~Bar, D.~Blas, K.~Blum, and S.~Sibiryakov, ``{Galactic rotation curves versus
  ultralight dark matter: Implications of the soliton-host halo relation},''
  \href{http://dx.doi.org/10.1103/PhysRevD.98.083027}{{\em Phys. Rev.}
  {\bfseries D98} no.~8, (2018) 083027},
\href{http://arxiv.org/abs/1805.00122}{{\ttfamily arXiv:1805.00122
  [astro-ph.CO]}}.

\bibitem{Safarzadeh:2019sre}
M.~Safarzadeh and D.~N. Spergel, ``{Ultra-light Dark Matter is Incompatible
  with the Milky Way's Dwarf Satellites},''
\href{http://arxiv.org/abs/1906.11848}{{\ttfamily arXiv:1906.11848
  [astro-ph.CO]}}.

\bibitem{Tremaine:1979we}
S.~Tremaine and J.~E. Gunn, ``{Dynamical Role of Light Neutral Leptons in
  Cosmology},'' \href{http://dx.doi.org/10.1103/PhysRevLett.42.407}{{\em Phys.
  Rev. Lett.} {\bfseries 42} (1979) 407--410}.
[,66(1979)].

\bibitem{ranjit2016zeptonewton}
G.~Ranjit, M.~Cunningham, K.~Casey, and A.~A. Geraci, ``Zeptonewton force
  sensing with nanospheres in an optical lattice,'' {\em Physical Review A}
  {\bfseries 93} no.~5, (2016) 053801.

\bibitem{myeong2017halo}
G.~Myeong, N.~Evans, V.~Belokurov, N.~Amorisco, and S.~Koposov, ``Halo
  substructure in the sdss--gaia catalogue: streams and clumps,'' {\em Monthly
  Notices of the Royal Astronomical Society} {\bfseries 475} no.~2, (2017)
  1537--1548.

\bibitem{o2018dark}
C.~A. O'Hare, C.~McCabe, N.~W. Evans, G.~Myeong, and V.~Belokurov, ``Dark
  matter hurricane: Measuring the s1 stream with dark matter detectors,'' {\em
  Physical Review D} {\bfseries 98} no.~10, (2018) 103006.

\bibitem{necib2018under}
L.~Necib, M.~Lisanti, S.~Garrison-Kimmel, A.~Wetzel, R.~Sanderson, P.~F.
  Hopkins, C.-A. Faucher-Gigu{\`e}re, and D.~Kere{\v{s}}, ``Under the
  firelight: Stellar tracers of the local dark matter velocity distribution in
  the milky way,'' \href{http://arxiv.org/abs/1810.12301}{{\ttfamily
  arXiv:1810.12301}}.

\bibitem{clerk2010introduction}
A.~A. Clerk, M.~H. Devoret, S.~M. Girvin, F.~Marquardt, and R.~J. Schoelkopf,
  ``Introduction to quantum noise, measurement, and amplification,'' {\em
  Reviews of Modern Physics} {\bfseries 82} no.~2, (2010) 1155.

\bibitem{blencowe2004quantum}
M.~Blencowe, ``Quantum electromechanical systems,'' {\em Physics Reports}
  {\bfseries 395} no.~3, (2004) 159--222.

\bibitem{aspelmeyer2014cavity}
M.~Aspelmeyer, T.~J. Kippenberg, and F.~Marquardt, ``Cavity optomechanics,''
  {\em Reviews of Modern Physics} {\bfseries 86} no.~4, (2014) 1391.

\bibitem{sheard2004observation}
B.~S. Sheard, M.~B. Gray, C.~M. Mow-Lowry, D.~E. McClelland, and S.~E.
  Whitcomb, ``Observation and characterization of an optical spring,'' {\em
  Physical Review A} {\bfseries 69} no.~5, (2004) 051801.

\bibitem{caves1981quantum}
C.~M. Caves, ``Quantum-mechanical noise in an interferometer,'' {\em Physical
  Review D} {\bfseries 23} no.~8, (1981) 1693.

\bibitem{aasi2013enhanced}
J.~Aasi, J.~Abadie, B.~Abbott, R.~Abbott, T.~Abbott, M.~Abernathy, C.~Adams,
  T.~Adams, P.~Addesso, R.~Adhikari, {\em et~al.}, ``Enhanced sensitivity of
  the ligo gravitational wave detector by using squeezed states of light,''
  {\em Nature Photonics} {\bfseries 7} no.~8, (2013) 613.

\bibitem{purdy2013strong}
T.~P. Purdy, P.-L. Yu, R.~W. Peterson, N.~S. Kampel, and C.~A. Regal, ``Strong
  optomechanical squeezing of light,'' {\em Physical Review X} {\bfseries 3}
  no.~3, (2013) 031012.

\bibitem{clark2017sideband}
J.~B. Clark, F.~Lecocq, R.~W. Simmonds, J.~Aumentado, and J.~D. Teufel,
  ``Sideband cooling beyond the quantum backaction limit with squeezed light,''
  {\em Nature} {\bfseries 541} no.~7636, (2017) 191.

\bibitem{braginsky1980quantum}
V.~B. Braginsky, Y.~I. Vorontsov, and K.~S. Thorne, ``Quantum nondemolition
  measurements,'' {\em Science} {\bfseries 209} no.~4456, (1980) 547--557.

\bibitem{BRAGINSKY1990251}
V.~Braginsky and F.~Khalili, ``Gravitational wave antenna with qnd speed
  meter,'' {\em Physics Letters A} {\bfseries 147} no.~5, (1990) 251 -- 256.

\bibitem{clerk2008back}
A.~Clerk, F.~Marquardt, and K.~Jacobs, ``Back-action evasion and squeezing of a
  mechanical resonator using a cavity detector,'' {\em New Journal of Physics}
  {\bfseries 10} no.~9, (2008) 095010.

\bibitem{hertzberg2010back}
J.~Hertzberg, T.~Rocheleau, T.~Ndukum, M.~Savva, A.~Clerk, and K.~Schwab,
  ``Back-action-evading measurements of nanomechanical motion,'' {\em Nature
  Physics} {\bfseries 6} no.~3, (2010) 213.

\bibitem{khosla2017quantum}
K.~E. Khosla, G.~A. Brawley, M.~R. Vanner, and W.~P. Bowen, ``Quantum
  optomechanics beyond the quantum coherent oscillation regime,'' {\em Optica}
  {\bfseries 4} no.~11, (2017) 1382--1387.

\bibitem{danilishin2018new}
S.~L. Danilishin, E.~Knyazev, N.~V. Voronchev, F.~Y. Khalili, C.~Gr{\"a}f,
  S.~Steinlechner, J.-S. Hennig, and S.~Hild, ``A new quantum speed-meter
  interferometer: measuring speed to search for intermediate mass black
  holes,'' {\em Light: Science \& Applications} {\bfseries 7} no.~1, (2018) 11.

\end{thebibliography}\endgroup

\end{document}